\documentclass[Physsubmission, Phys]{SciPost}

\hypersetup{
    colorlinks,
    linkcolor={red!50!black},
    citecolor={blue!50!black},
    urlcolor={blue!80!black}
}

\usepackage[bitstream-charter]{mathdesign}
\DeclareSymbolFont{usualmathcal}{OMS}{cmsy}{m}{n}
\DeclareSymbolFontAlphabet{\mathcal}{usualmathcal}

\begin{document}

\begin{center}{\Large \textbf{
Top quark mass corrections to single and double
Higgs boson production in gluon fusion\\
}}\end{center}

\begin{center}
J. Davies\textsuperscript{1*}
\end{center}

\begin{center}
{\bf 1} Department of Physics and Astronomy, University of Sussex, Brighton, BN1 9HQ, UK 
\\
* j.o.davies@sussex.ac.uk
\end{center}

\begin{center}
\today
\end{center}


\definecolor{palegray}{gray}{0.95}
\begin{center}
\colorbox{palegray}{
  \begin{tabular}{rr}
  \begin{minipage}{0.1\textwidth}
    \includegraphics[width=35mm]{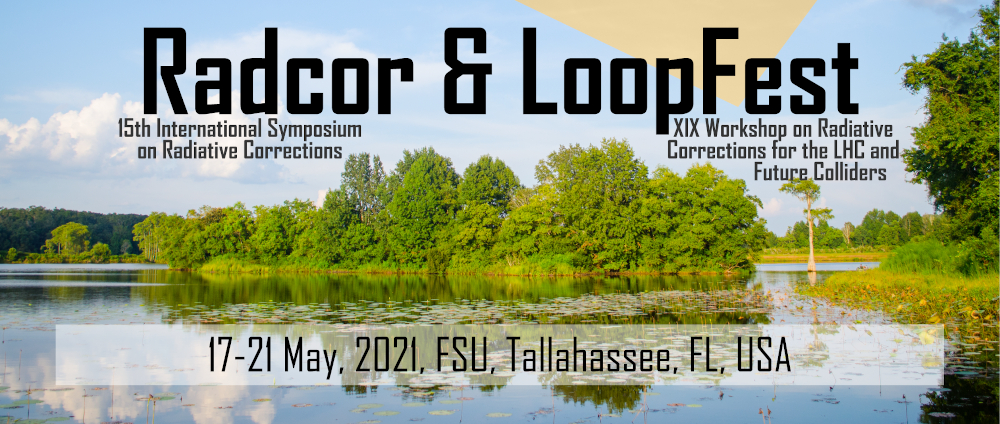}
  \end{minipage}
  &
  \begin{minipage}{0.85\textwidth}
    \begin{center}
    {\it 15th International Symposium on Radiative Corrections: \\Applications of Quantum Field Theory to Phenomenology,}\\
    {\it FSU, Tallahasse, FL, USA, 17-21 May 2021} \\
    \doi{10.21468/SciPostPhysProc.?}\\
    \end{center}
  \end{minipage}
\end{tabular}
}
\end{center}


\definecolor{proc-orange}{HTML}{800000}

\newcommand{\as}{\ensuremath{\alpha_s}}
\newcommand{\tf}{\ensuremath{T_F}}
\newcommand{\mt}{\ensuremath{m_t}}
\newcommand{\mts}{\ensuremath{m_t^{\,2}}}
\newcommand{\mh}{\ensuremath{m_H}}
\newcommand{\mhs}{\ensuremath{m_H^{\,2}}}
\newcommand{\mhc}{\ensuremath{m_H^{\,3}}}


\section*{Abstract}
{\bf
In this talk we discuss recent computations of the top quark mass dependence of QCD amplitudes
describing Higgs boson production in gluon fusion. We compute terms in the expansion
for a large top quark mass, which reduces the Feynman diagrams to products of massless
integrals and massive tadpole integrals which contain the top mass dependence. In
particular we discuss the real and virtual corrections to double Higgs production at NNLO,
and the virtual corrections to single Higgs production at N3LO.
}



\section{Introduction}
\label{sec:intro}
One of the numerous tasks of the Large Hadron Collider (LHC) is to characterize the
structure of the Standard Model's (SM) scalar sector. The parameter $\lambda$ governs
triple and quartic Higgs boson interactions, and is determined in the SM by the mass and
vacuum expectation value of the Higgs boson. A direct experimental measurement of
$\lambda$ will help determine if the SM's scalar sector is observed in nature, although
such a measurement is very challenging~\cite{ATLAS:2020jgy,CMS:2020tkr}.

It is therefore important to have a good theoretical understanding of processes which
involve a single Higgs boson and two Higgs bosons.
Such processes tend to be dominated by contributions
with top quarks propagating in loops, due to the large value of the top quark's
Yukawa coupling. Multi-loop amplitudes quickly reach a complexity which cannot currently
be handled in an exact manner, so we turn to approximation methods in order to study them.
In particular, here, we discuss expansions which consider the top quark mass to be larger
than any other scale involved in the problem. The results of such expansions,
in addition to providing a good description of amplitudes below the top quark threshold,
can be combined with information describing other kinematic regions, to produce
approximations which describe amplitudes over a wider kinematic range, see for
e.g.~Refs.~\cite{Grober:2017uho,Davies:2019nhm,Czakon:2020vql}.
A description of the large-$\mt$ expansion method is given in
Section~\ref{sec:LME} and of some of its applications in Section~\ref{sec:appl}.


\section{Large Mass Expansion (LME)}
\label{sec:LME}

The comparatively large value of the mass of the top quark means that performing an
expansion in the limit $\mt \to \infty$ leads to a sensible approximation of scattering
amplitudes, particularly for processes in which contributions due to top quarks dominate.
Amplitudes involving Higgs bosons are examples of such processes; the size of the top
quark Yukawa coupling means that contributions from other quark flavours are relatively
unimportant. The leading term in such an expansion yields the so-called ``Higgs effective
field theory'' (HEFT), in which the top quark is completely integrated out. Dependence on
its mass appears only logarithmically in the effective couplings of Higgs bosons and
gluons. The goal of the computations discussed in these proceedings is to include
\emph{sub-leading} terms of the LME, i.e., terms proportional to powers of $1/\mts$.

\begin{figure}
	\begin{center}
	\begin{tabular}{ccc}
		\hline
		Full graph & Subgraph & Co-subgraph $\times$ Expanded subgraph\\
		\hline
		\includegraphics[height=2cm]{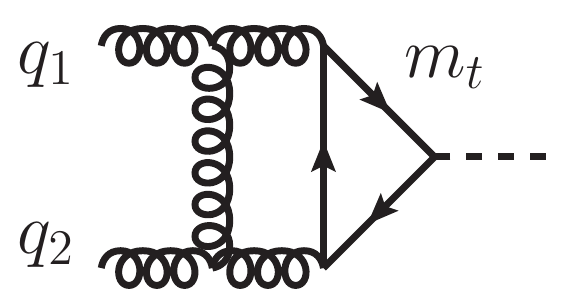} &
		\includegraphics[height=2cm]{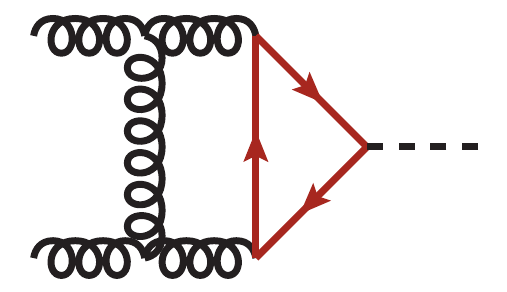} &
		\includegraphics[height=2cm]{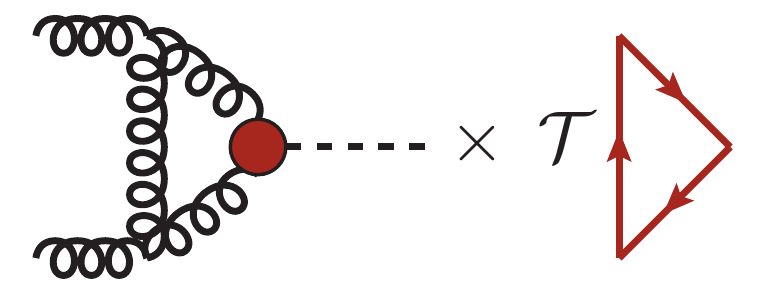}
		\\ 
		\cline{2-3}
		&
		\includegraphics[height=2cm]{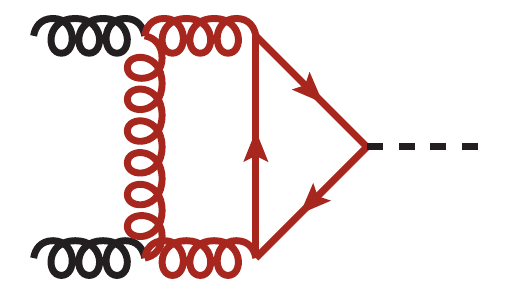} &
		\includegraphics[height=2cm]{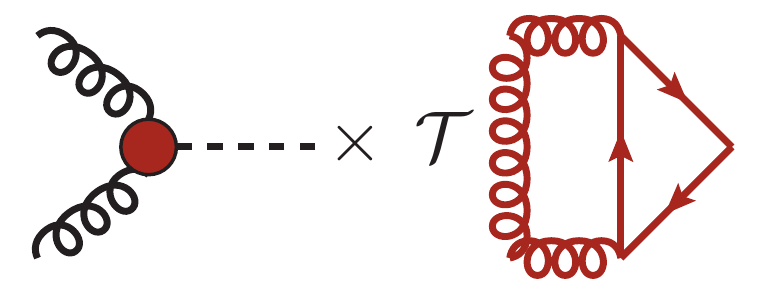}
		\\ 
		\hline 	
	\end{tabular}
	\caption{\label{fig:LMEex}\emph{
		Starting from the full Feynman graph of the first column, we identify the
		subgraphs which contain $\mt$-dependent propagators. To perform the LME, the
		subgraphs are expanded in the large-$\mt$ limit (denoted by the operator
		$\mathcal{T}$), leaving only co-subgraphs which are independent of $\mt$.
	}}
	\end{center}
\end{figure}

Such an asymptotic expansion can be performed by the method of ``expansion by subgraph'',
which is conveniently implemented in the program \texttt{exp}~\cite{Harlander:1998cmq,
Seidensticker:1999bb}. The procedure is to identify all subgraphs of a given Feynman
graph which contain the heavy scale (here, $\mt$) and expand them in their small quantities.
The remaining propagators form the ``co-subgraph'' which does not depend on the heavy scale.
For the problem at hand, this means that each Feynman graph is reduced to a sum of products
of massless graphs and $\mt$-dependent vacuum graphs, after the expansion in the limit
$\mt \to \infty$. This procedure is depicted in Fig.~\ref{fig:LMEex}. For the subgraph
identified on the first row, the propagators are expanded as follows:
\begin{flalign}
	\int\!\!\!&\int \mbox{d}^d l_1 \: \mbox{d}^d l_2 \:
		\frac{1}{l_2^2}
		\frac{1}{(l_2+q_1)^2}
		\frac{1}{(l_2-q_2)^2}
		{\color{proc-orange}
			\frac{1}{(l_1+q_1)^2-\mts}
			\frac{1}{(l_1-q_2)^2-\mts}
			\frac{1}{(l_1-l_2)^2-\mts}
		}
		\longrightarrow
\nonumber\\
	& \:\:\:\: \int \mbox{d}^d l_2 \:
		\frac{1}{l_2^2}
		\frac{1}{(l_2+q_1)^2}
		\frac{1}{(l_2-q_2)^2}
			\int \mbox{d}^d l_1\!
			\left[
				\frac{1}{(l_1^2-\mts)^3}
				+ \frac{2(q_1\!\cdot\!l_1 - q_2\!\cdot\!l_1 - l_2\!\cdot\!l_1) + l_2\!\cdot\!l_2}{(l_1^2-\mts)^4}
				+ \cdots
			\right]\,,
\end{flalign}
where the ``$\cdots$'' represent higher-order terms in the expansion. Within the square
brackets, the propagators do not carry external momenta $q_1$ or $q_2$, nor the loop momentum $l_2$;
there is only a sum of one-loop vacuum integrals with tensor numerators. Such integrals
can be computed by the \texttt{FORM}~\cite{Ruijl:2017dtg} package
\texttt{MATAD}~\cite{Steinhauser:2000ry}, up to three-loop order.
For computations which require four-loop vacuum integrals we make use of \texttt{FIRE}~\cite{Smirnov:2019qkx},
which we also use to perform integration-by-parts reduction for the massless co-subgraphs.
A more complete description of the computational toolchain is given in the following section.

\subsection{Computational Toolchain}

For all computations, we begin by generating the necessary Feynman diagrams with
\texttt{qgraf}~\cite{Nogueira:1991ex}. From here, the packages \texttt{q2e} and
\texttt{exp}~\cite{Harlander:1998cmq,Seidensticker:1999bb} are used
to convert the diagrams into a compatible notation and to identify the relevant subgraphs
and co-subgraphs, as described in Section~\ref{sec:LME}. Code is generated for the expansion
to be performed by \texttt{FORM}~\cite{Ruijl:2017dtg}, which also computes the colour factors
using the \texttt{COLOR} package~\cite{vanRitbergen:1998pn}.
Vacuum graphs up to three loops are computed with
the \texttt{MATAD} package~\cite{Steinhauser:2000ry}, and four-loop vacuum graphs as well
as the massless
integrals of the co-subgraphs are reduced to master integrals using
\texttt{FIRE}~\cite{Smirnov:2019qkx}.
For the computation of phase-space integrals described in Section~\ref{sec:ggHHreal}, we use
\texttt{LiteRed}~\cite{Lee:2012cn,Lee:2013mka} and \texttt{LIMIT}~\cite{Herren:2020ccq}.


\section{Applications}
\label{sec:appl}
In the following, we summarize some recent works which have used the LME method
described above. This includes an NNLO computation of the double-real and
real-virtual corrections to double--Higgs boson production in gluon
fusion~\cite{Davies:2019esq,Davies:2019xzc,Davies:2021kex}
in Section~\ref{sec:ggHHreal},
and N3LO computations of the virtual corrections to single--Higgs boson
production in gluon fusion~\cite{Davies:2019wmk} in Section~\ref{sec:ggH4l}, as well as the
decay of a Higgs boson into two photons~\cite{Davies:2021zbx} in Section~\ref{sec:Haa4l}.

\subsection{NNLO real-radiation corrections to double Higgs boson production}
\label{sec:ggHHreal}
\begin{figure}
	\centering
	\includegraphics[height=3cm]{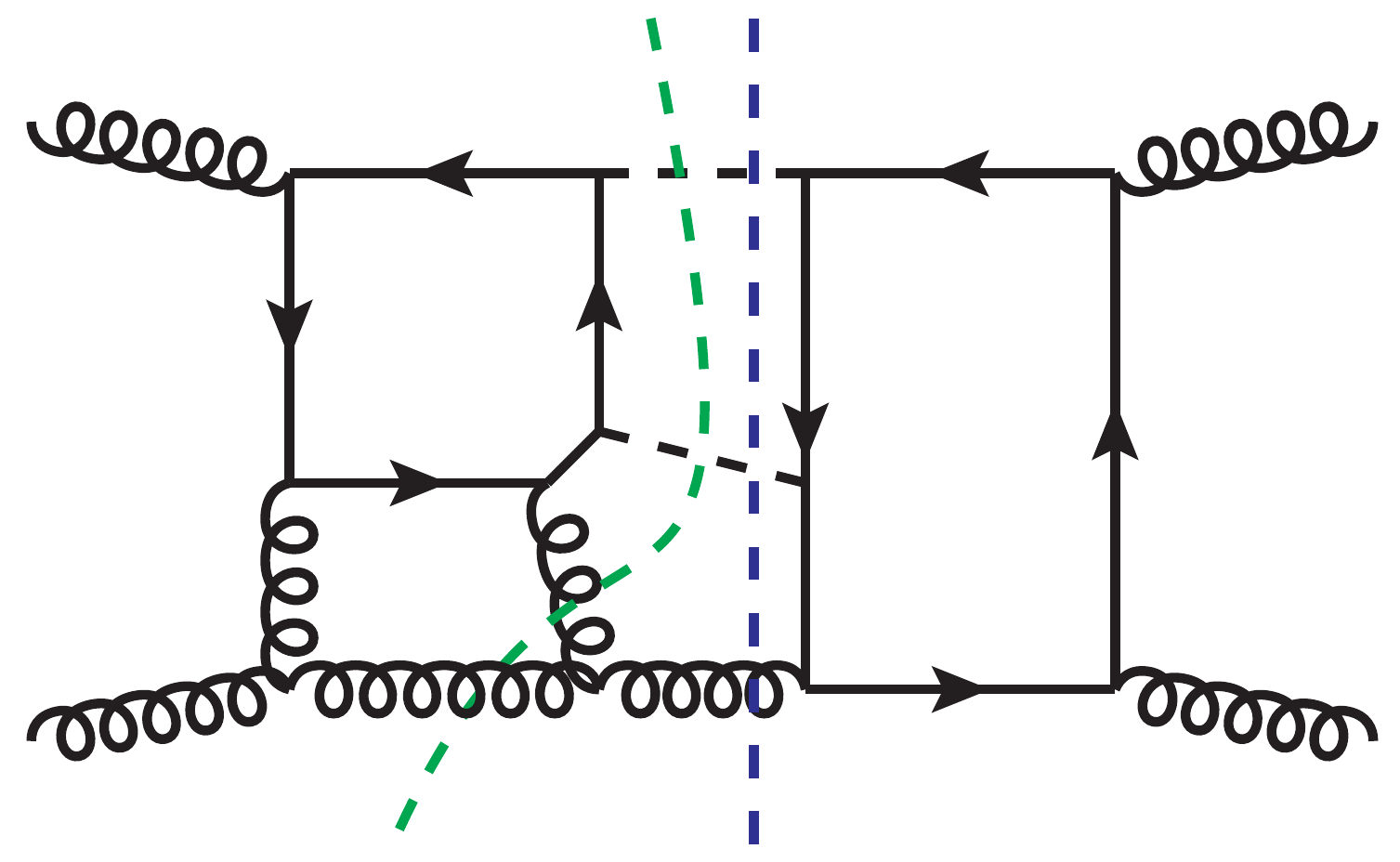}
	\includegraphics[height=3cm]{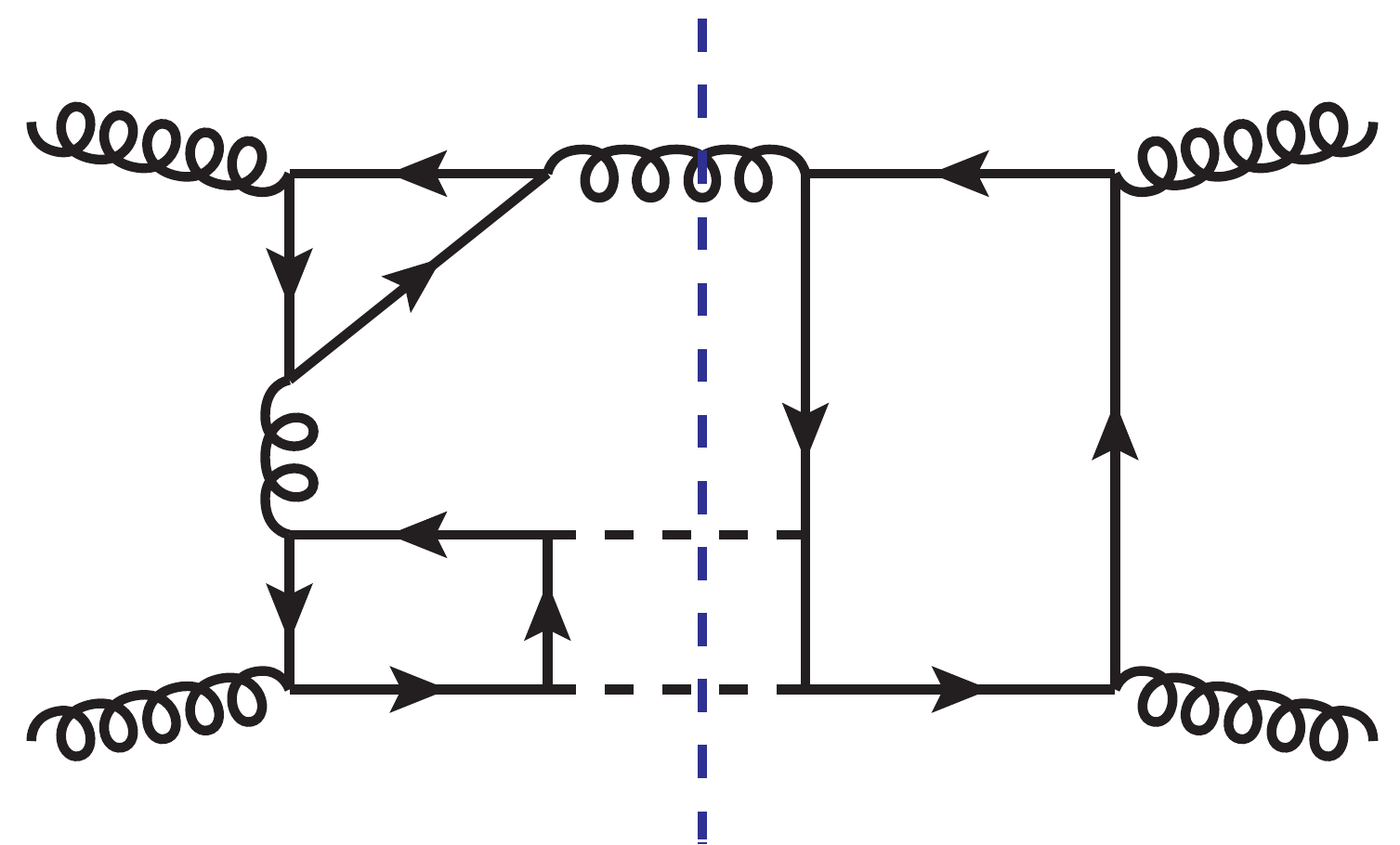}
	\includegraphics[height=3cm]{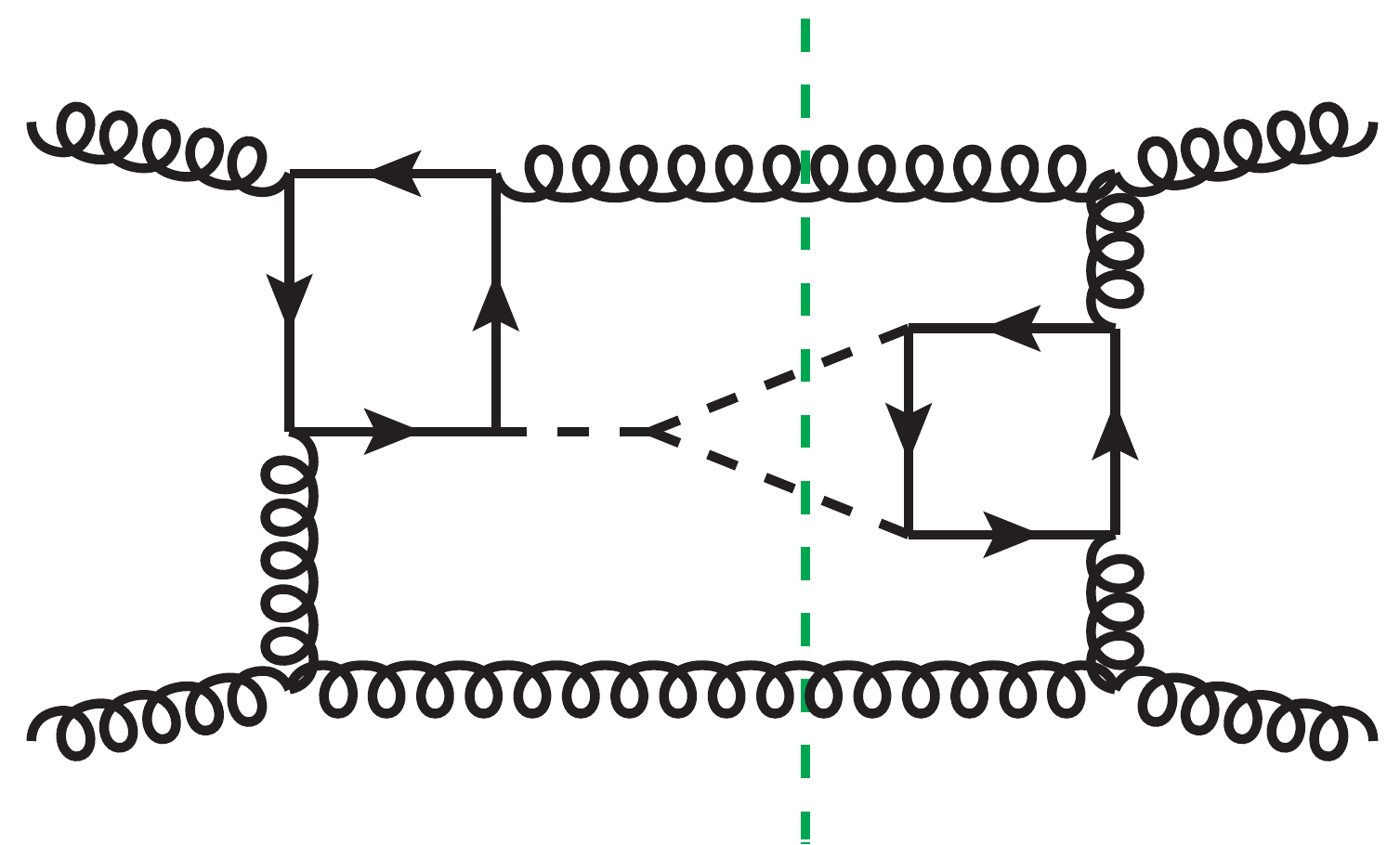}
	\caption{\emph{Forward-scattering diagrams which contribute to the real-real and real-virtual corrections to NNLO double--Higgs boson production. The blue and green dashed lines denote three- and four-particle cuts through the final state particles.}}
	\label{fig:forward-scattering}
\end{figure}

In order to compute real-real and real-virtual corrections we make use
of the optical theorem, and compute forward-scattering diagrams which
have cuts through the desired final state particles. The real-real
corrections correspond to cuts through two Higgs bosons and two additional
particles, and the real-virtual corrections to cuts through two Higgs bosons and one additional
particle. Examples of such cuts are given in Fig.~\ref{fig:forward-scattering}, where the
three-particle cuts are shown by blue dashed lines and four-particle cuts by green dashed lines.
Some diagrams, such as the first, admit both a three- and four-particle cut. To generate such
diagrams an additional step is required to post-process the output of \texttt{qgraf}, which can
not itself generate only diagrams containing specified cuts. For this purpose we use the program
\texttt{gen}\footnote{A.~Pak, Unpublished.}.

\begin{figure}
	\centering
	\includegraphics[height=2.9cm]{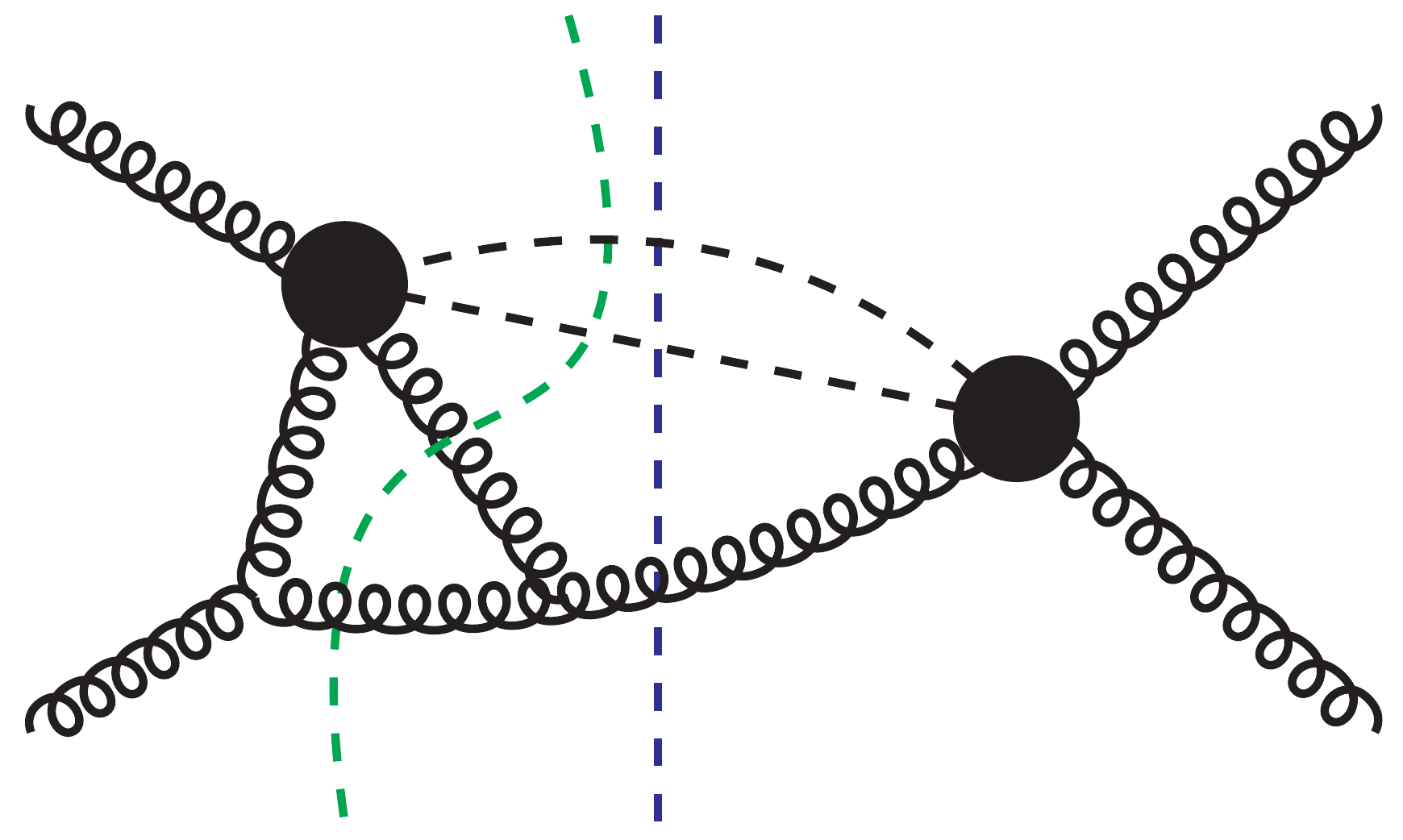}
	\includegraphics[height=2.9cm]{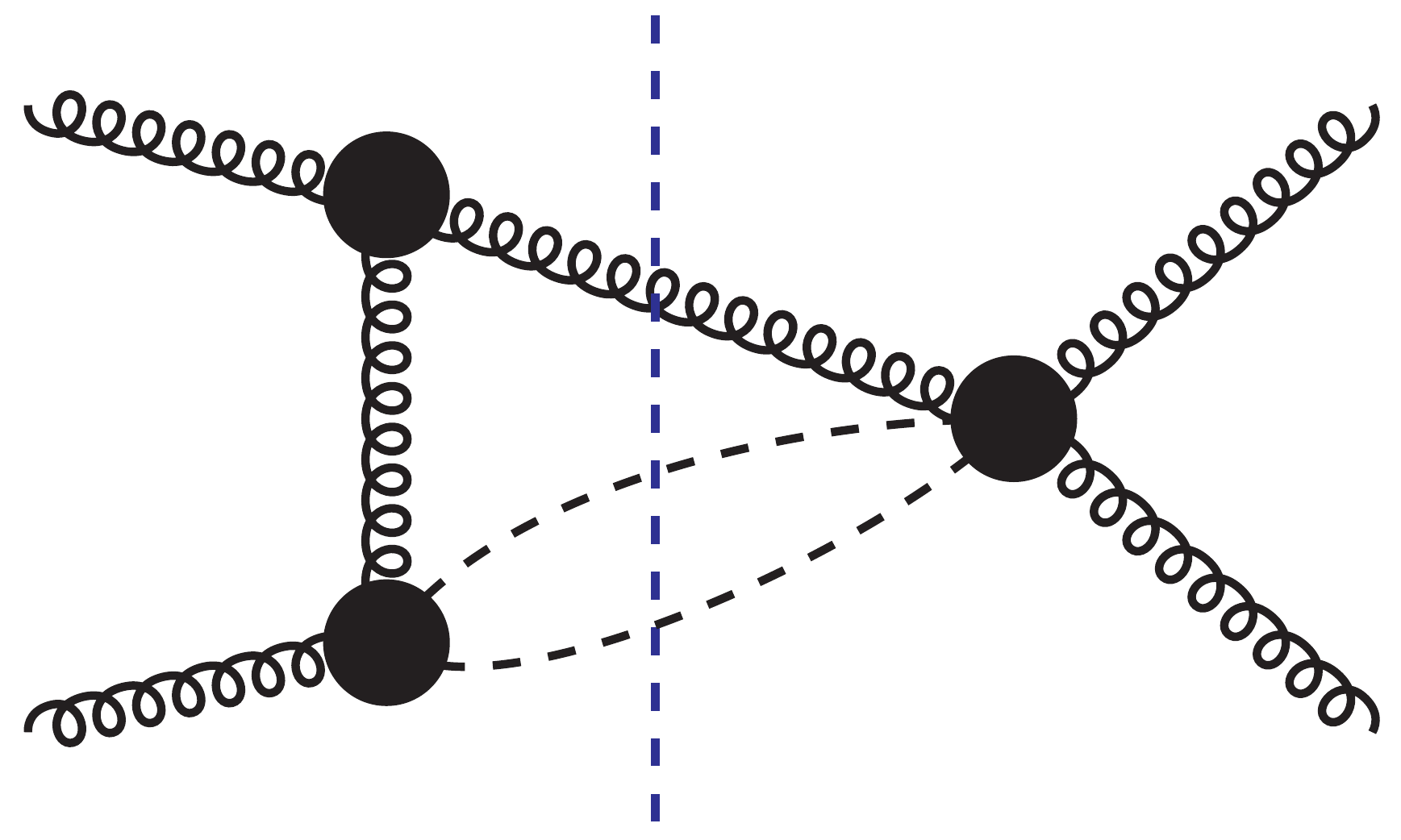}
	\includegraphics[height=2.9cm]{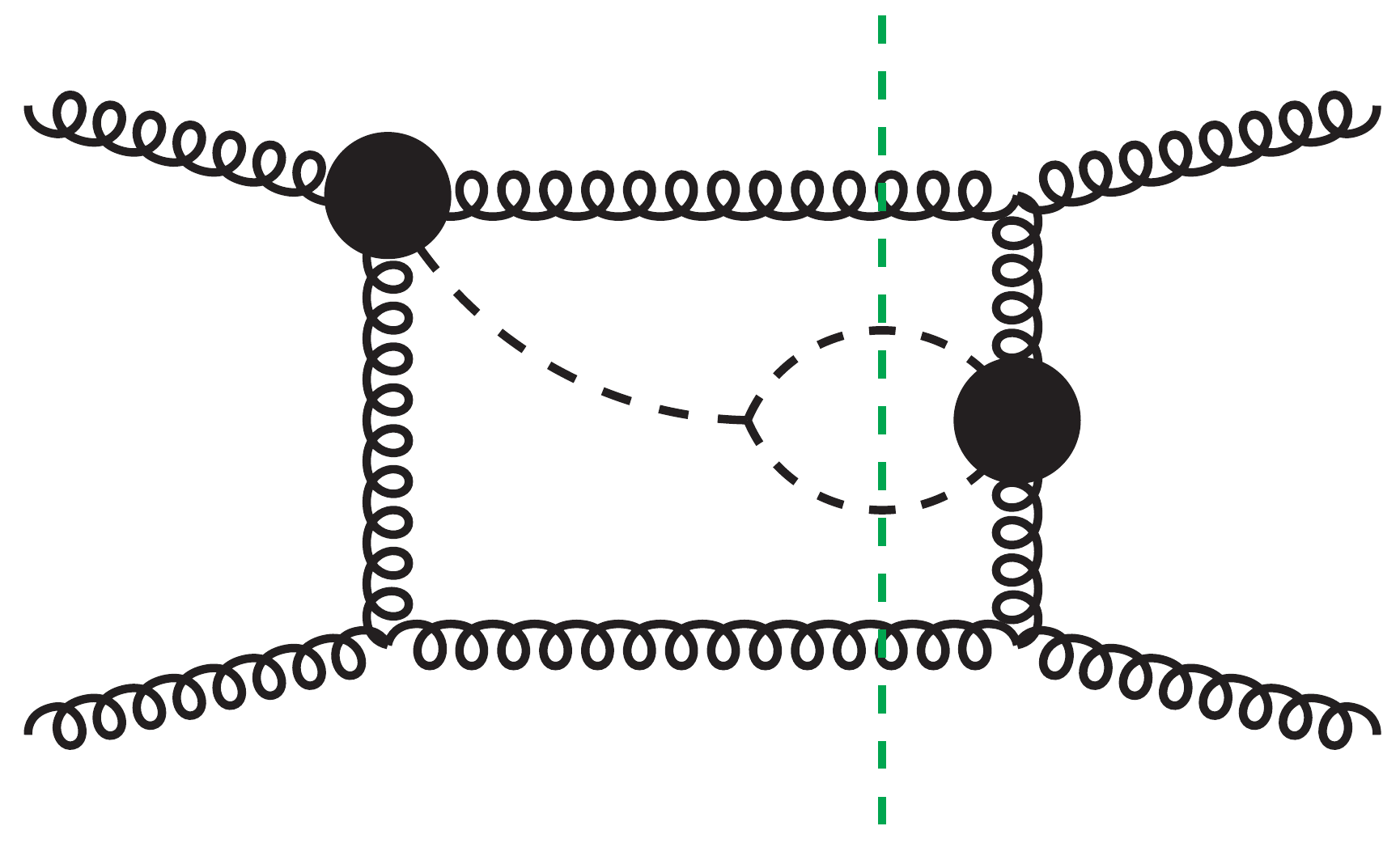}
	\caption{\emph{Forward-scattering diagrams which contribute to the real-real and real-virtual corrections to NNLO double--Higgs boson production. The blue and green dashed lines denote three- and four-particle cuts through the final state particles.}}
	\label{fig:forward-scattering-lme}
\end{figure}

After large-$\mt$ expansion, the diagrams of Fig.~\ref{fig:forward-scattering} yield the phase-space
integrals shown in Fig.~\ref{fig:forward-scattering-lme}. These integrals are reduced to a basis of
master integrals by \texttt{LiteRed} after the partial fractioning of linearly-dependent propagators
by \texttt{LIMIT}.
The master integrals are computed in an expansion around $\delta = 1-4\mhs/s \to 0$, which corresponds
to the production threshold of the Higgs boson pair. A deep expansion in $\delta$ is obtained
efficiently through the use of differential equations, starting from boundary values computed
for the leading term in the expansion.

The total cross section at NNLO is given by
\begin{equation}
	\label{eq:totalXS}
	\sigma_{ij}^{(2)} =
		\sigma_{ij,\mathrm{real}}^{(2)}
		+ \sigma_{ij,\mathrm{coll}}^{(2)}
		+ \sigma_{ij,\mathrm{virt}}^{(2)}\,,
\end{equation}
where $i,j$ denote the contributing partonic channels, $gg$, $gq$, $q\bar{q}$, $qq$, $qq'$, including
also all additional anti-quark and ghost channels.
The collinear counterterms are computed through the convolution of the LO and NLO cross sections
with quark or gluon splitting functions.
See Section~2.5 of Ref.~\cite{Davies:2021kex} for a detailed discussion. The
virtual corrections have been computed in Ref.~\cite{Grigo:2015dia,Davies:2019djw}.
After ultra-violet renormalization, the sum of
contributions in Eq.~(\ref{eq:totalXS}) produces a finite result.

\begin{figure}
	\centering
	\hspace{0mm}\includegraphics[height=3.8cm]{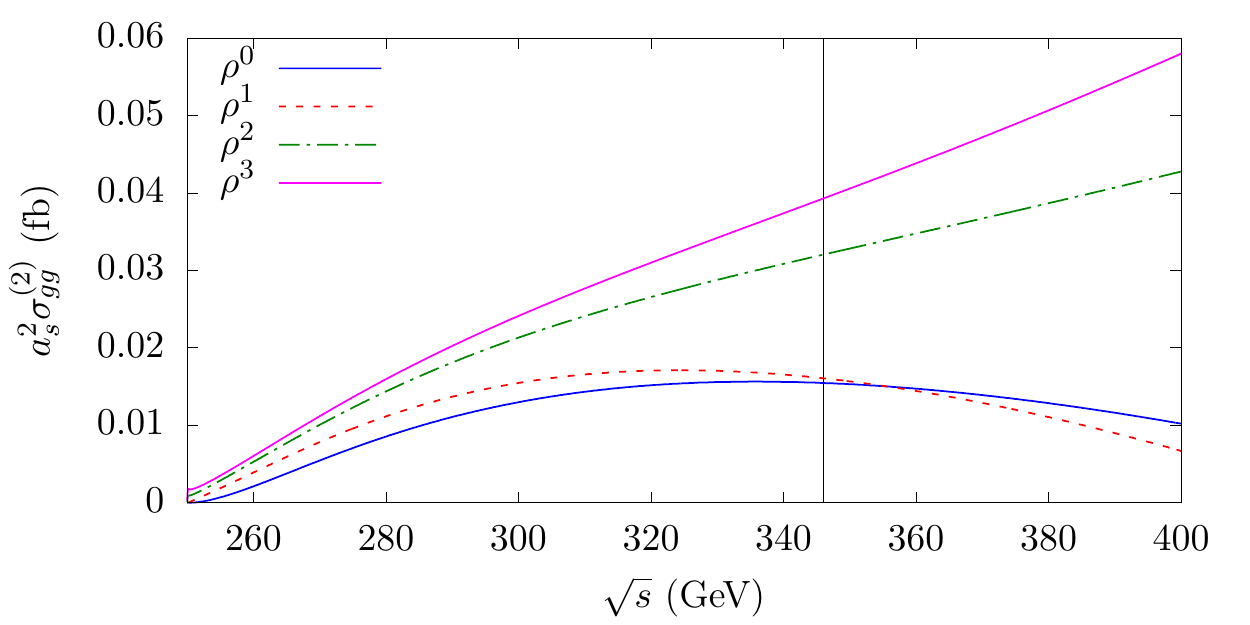}
	\hspace{-3mm}\includegraphics[height=3.8cm]{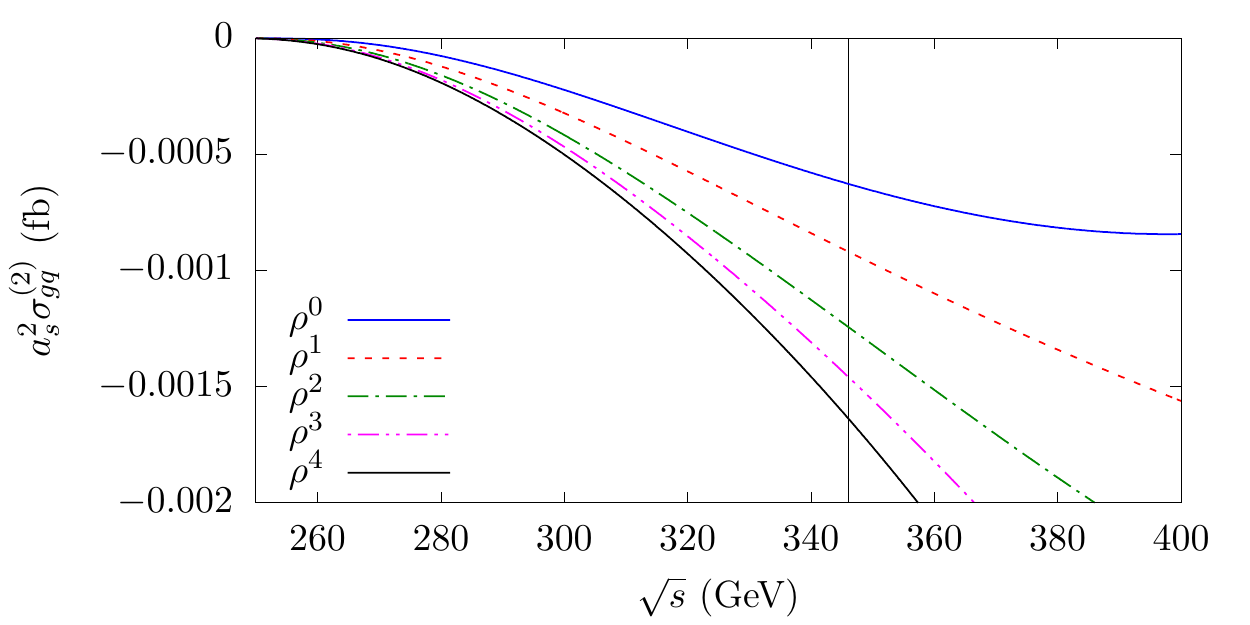}
	\caption{\emph{NNLO contributions to the total cross sections of the $gg$ and $gq$ channels.
		The vertical black line shows the threshold at $\sqrt{s}=2\mt$. The curves show different
		orders in the large-$\mt$ expansion, denoted by $\rho^n = (\mhs/\mts)^n$.}}
	\label{fig:totalXSresults}
\end{figure}

Plots for successive orders of the
large-$\mt$ expansion are shown, for the $gg$ and $gq$ channels, in Fig.~\ref{fig:totalXSresults}.
Below the top quark threshold the large-$\mt$ expansion converges well, particularly so close to
the production threshold.
Compared to the leading expansion term ($\rho^0$), which corresponds to the HEFT result (which
has been previously computed in Ref.~\cite{deFlorian:2013jea}), including
the sub-leading terms typically corrects the NNLO contributions by a factor of two.

\subsection{N3LO virtual corrections to single Higgs boson production}
\label{sec:ggH4l}
\begin{figure}
	\centering
	\includegraphics[height=2cm]{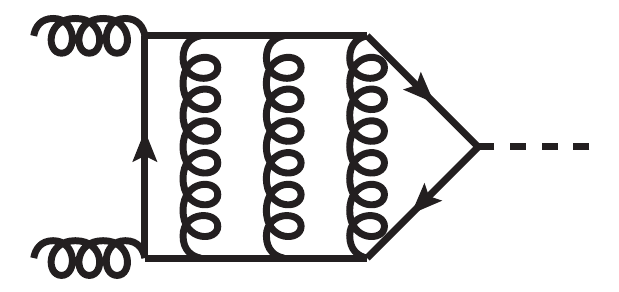}
	\includegraphics[height=2cm]{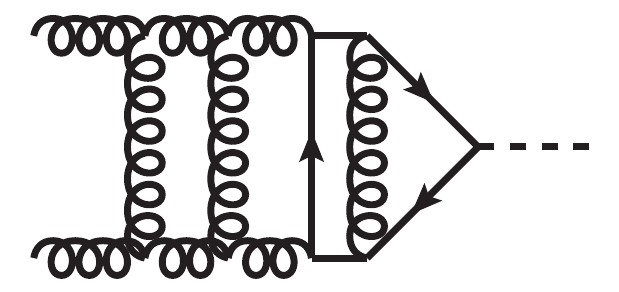}
	\caption{\emph{Virtual corrections contributing to $gg\to H$ at N3LO.}}
	\label{fig:ggH4l}
\end{figure}

For $2\to 1$ processes, it is computationally feasible to compute virtual corrections at
N3LO, corresponding to four-loop order for loop-induced processes such as single--Higgs boson
production in gluon fusion, and Higgs boson decay into two photons (see Section~\ref{sec:Haa4l}).
Here we compute diagrams such as those shown in Fig.~\ref{fig:ggH4l}. The expansion proceeds in a
straightforward manner, as outlined in Section~\ref{sec:LME}.
We define the amplitude to be
\begin{equation}
	\mathcal{A} = \frac{4\as(\mu)}{3\pi}\frac{\tf}{\nu}
		\delta^{ab}
		\left(q_1\!\cdot\!q_2\,g_{\vphantom{12}}^{\mu\nu}
			- q_1^{\,\nu\vphantom{\mu}}q_2^{\,\mu}\right)
		\:h(\rho)\,,
\end{equation}
where $q_1, q_2$ are the momenta of the incoming gluons, $a, b$ their colour indices, $\tf=1/2$,
and $\nu$ is the Higgs vacuum expectation value. After ultra-violet renormalization, the form factor
$h(\rho)$ still contains infra-red poles, however the poles of the rescaled form factor
\begin{equation}
	F(\rho) = h(\rho)/h^{(1)}(\rho) = 1 + \mathcal{O}(\as)
\end{equation}
are predicted to factorize and given in the literature~\cite{Gehrmann:2010ue}. We thus consider
\begin{equation}
	\log{(F)} = \log{(F)}_{\mathrm{poles}} + \log{(F)}_{\mathrm{finite}}
\end{equation}
and find that indeed $\log{(F)}_{\mathrm{poles}}$ is independent of $\rho$ and agrees
with Ref.~\cite{Gehrmann:2010ue},
and we may study the numerical impact of the N3LO corrections to $\log{(F)}_{\mathrm{finite}}$.
At a renormalization scale $\mu=\mt$ with an on-shell $\mt$ value of 173~GeV, we find that
\begin{align}
	\log{(F)}_{\mathrm{finite}} =
		& {} + a_t^{\phantom{2}} \left[
			{} + (11.07-3.06i) + (0.07) + (0.004)
			\right]
		\nonumber\\
		& {} + a_t^{2} \left[
			{} + (22.59-13.24i) + (1.02-0.13i) + (0.07-0.01i)
			\right]
		\nonumber\\
		& {} + a_t^{3} \left[
			{} - (73.18+51.55i) + (7.61+0.85i) + (0.70-0.14i)
			\right]\,,
\end{align}
where the $\rho^0$, $\rho^1$ and $\rho^2$ expansion terms have been displayed individually in
round brackets.
We observe that the sub-leading terms in the large-$\mt$ expansion become increasingly important
at higher perturbative orders; at N3LO the $\rho^1$ and $\rho^2$ terms correct the $\rho^0$ term
by about 10\% and 1\%, respectively.


\subsection{N3LO corrections to Higgs boson decay into photons}
\label{sec:Haa4l}
From a computational point of view, the decay process $H\to\gamma\gamma$ is very similar
to $gg\to H$ discussed above in Section~\ref{sec:ggH4l}. The Feynman diagrams which contribute
are a subset of those of $gg\to H$, i.e., those for which the external gluons (now photons)
couple to a quark line rather than internally propagating gluons. The first diagram of
Fig.~\ref{fig:ggH4l} is such an example. We can therefore apply our existing machinery and
reduction to master integrals to this process. We define the partial decay width as
\begin{equation}
	\Gamma_{H\to\gamma\gamma} = \frac{\mhc}{64\pi}\left|A(\rho)\right|,
\end{equation}
and compute $A(\rho)$ in the large-$\mt$ expansion. Unlike the form factor $h(\rho)$ of
$gg\to H$, $A(\rho)$ is finite after ultra-violet renormalization.

\begin{figure}
	\centering
	\includegraphics[height=5cm]{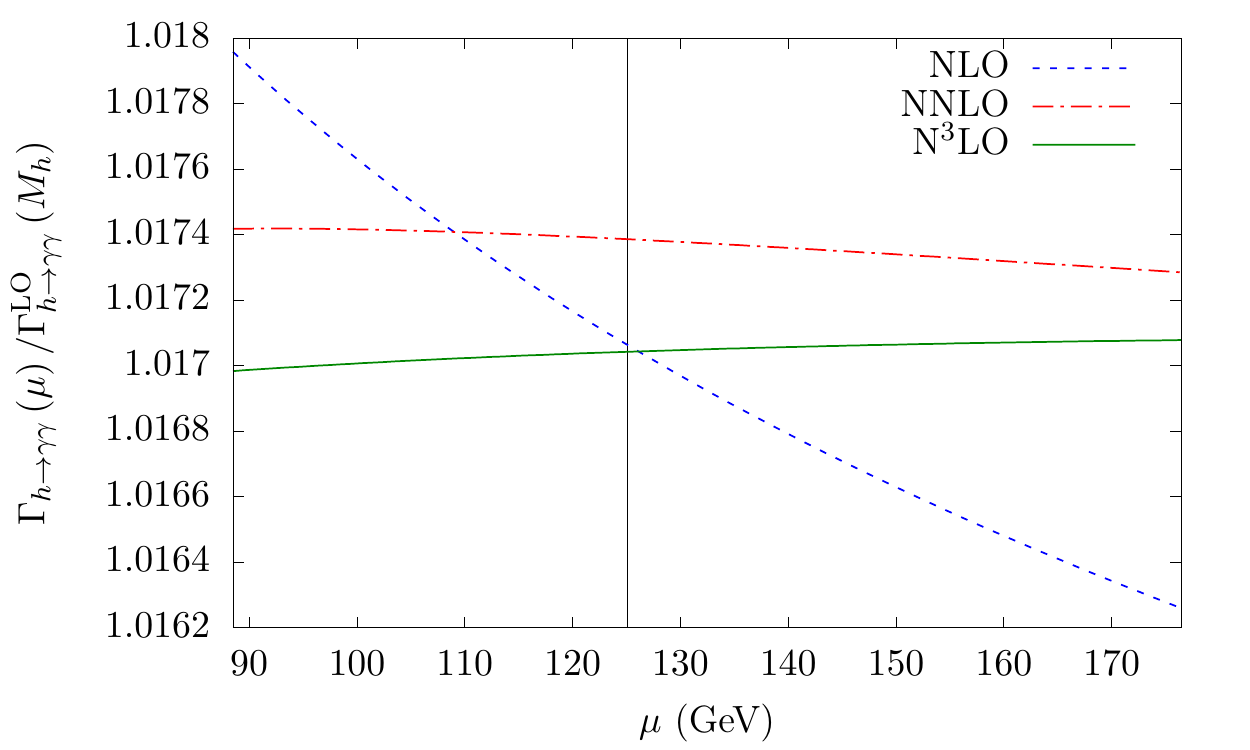}
	\caption{\emph{The scale dependence of the large-$\mt$ expansions of the NLO, NNLO and
		N3LO corrections to $\Gamma_{H\to\gamma\gamma}$, w.r.t.~the exact leading order.
		$\mt$ is in the on-shell scheme. The vertical black line is at $\mu=\mh$.}}
	\label{fig:Haa4lMuDepOS}
\end{figure}

For the top quark mass in the on-shell scheme, Fig.~\ref{fig:Haa4lMuDepOS} shows the dependence
on the renormalization scale of the NLO, NNLO and N3LO corrections, w.r.t.~the leading order.
At N3LO the curve becomes slightly flatter, but it does not overlap with the NNLO curve.

The N3LO large-$\mt$ result can be combined with the NLO electroweak
corrections~\cite{Actis:2008ts} as
well as the NNLO corrections due to bottom and charm quark loops~\cite{Niggetiedt:2020sbf},
\begin{align}
	\label{eq:Haa4ldecay}
	\Gamma_{H\to\gamma\gamma} \times 10^{6} \text{GeV}^{-1} &=
		9.2581|_{\text{LO}}
		- 0.1502|_{\text{NLO,EW}}
		+ 0.1569|_{\text{NLO,t}}
		+ 0.0157|_{\text{NLO,bc}}
		\nonumber\\
		&{} + 0.0029|_{\text{NNLO,t}}
		+ 0.0036|_{\text{NNLO,bc}}
		- 0.0031|_{\text{N$^3$LO,t}}
		= 9.2838\,.
\end{align}
We find that the N3LO top quark corrections are about the same size as each of the NNLO
corrections shown in Eq.~(\ref{eq:Haa4ldecay}), but come with the opposite sign.


\section{Conclusion}
In these proceedings, we have discussed the large-$\mt$ expansion and its application to
several scattering processes. For processes involving Higgs bosons, the size of the top
quark Yukawa coupling means that amplitudes with top quarks running in the loops contribute
an important part of the total cross sections. The expansion allows the effect of the top
quark mass to be well described below the threshold, and including sub-leading terms typically
produces large corrections w.r.t.~the leading term alone, which corresponds to the
commonly-used HEFT.

Such expansions can be performed in a semi-automated and systematic way, allowing us to study
multi-loop amplitudes which can not be computed in an exact manner, either analytically or
numerically. As discussed in Section~\ref{sec:intro}, these expansions form the input for
various approximation methods which combine information from various kinematic regions, in
an attempt to describe multi-loop amplitudes over a wider kinematic range than any one
expansion alone.

\section*{Acknowledgements}
The work of JD was partly supported by the Science and Technology Facilities Council (STFC) under the Consolidated Grant ST/T00102X/1.

\bibliography{radcor-proceedings.bib}

\begin{thebibliography}{10}
\providecommand{\url}[1]{\texttt{#1}}
\providecommand{\urlprefix}{URL }
\expandafter\ifx\csname urlstyle\endcsname\relax
  \providecommand{\doi}[1]{doi:\discretionary{}{}{}#1}\else
  \providecommand{\doi}{doi:\discretionary{}{}{}\begingroup
  \urlstyle{rm}\Url}\fi
\providecommand{\eprint}[2][]{\url{#2}}

\bibitem{ATLAS:2020jgy}
G.~Aad \emph{et~al.},
\newblock \emph{{Search for the $HH \rightarrow b \bar{b} b \bar{b}$ process
  via vector-boson fusion production using proton-proton collisions at
  $\sqrt{s} = 13$ TeV with the ATLAS detector}},
\newblock JHEP \textbf{07}, 108 (2020),
\newblock \doi{10.1007/JHEP07(2020)108},
\newblock [Erratum: JHEP 01, 145 (2021), Erratum: JHEP 05, 207 (2021)],
\newblock \eprint{2001.05178}.

\bibitem{CMS:2020tkr}
A.~M. Sirunyan \emph{et~al.},
\newblock \emph{{Search for nonresonant Higgs boson pair production in final
  states with two bottom quarks and two photons in proton-proton collisions at
  $ \sqrt{s} $ = 13 TeV}},
\newblock JHEP \textbf{03}, 257 (2021),
\newblock \doi{10.1007/JHEP03(2021)257},
\newblock \eprint{2011.12373}.

\bibitem{Grober:2017uho}
R.~Gr\"ober, A.~Maier and T.~Rauh,
\newblock \emph{{Reconstruction of top-quark mass effects in Higgs pair
  production and other gluon-fusion processes}},
\newblock JHEP \textbf{03}, 020 (2018),
\newblock \doi{10.1007/JHEP03(2018)020},
\newblock \eprint{1709.07799}.

\bibitem{Davies:2019nhm}
J.~Davies, R.~Gr\"ober, A.~Maier, T.~Rauh and M.~Steinhauser,
\newblock \emph{{Top quark mass dependence of the Higgs boson-gluon form factor
  at three loops}},
\newblock Phys. Rev. D \textbf{100}(3), 034017 (2019),
\newblock \doi{10.1103/PhysRevD.100.034017},
\newblock [Erratum: Phys.Rev.D 102, 059901 (2020)],
\newblock \eprint{1906.00982}.

\bibitem{Czakon:2020vql}
M.~L. Czakon and M.~Niggetiedt,
\newblock \emph{{Exact quark-mass dependence of the Higgs-gluon form factor at
  three loops in QCD}},
\newblock JHEP \textbf{05}, 149 (2020),
\newblock \doi{10.1007/JHEP05(2020)149},
\newblock \eprint{2001.03008}.

\bibitem{Harlander:1998cmq}
R.~Harlander, T.~Seidensticker and M.~Steinhauser,
\newblock \emph{{Complete corrections of Order alpha alpha-s to the decay of
  the Z boson into bottom quarks}},
\newblock Phys. Lett. B \textbf{426}, 125 (1998),
\newblock \doi{10.1016/S0370-2693(98)00220-2},
\newblock \eprint{hep-ph/9712228}.

\bibitem{Seidensticker:1999bb}
T.~Seidensticker,
\newblock \emph{{Automatic application of successive asymptotic expansions of
  Feynman diagrams}},
\newblock In \emph{{6th International Workshop on New Computing Techniques in
  Physics Research: Software Engineering, Artificial Intelligence Neural Nets,
  Genetic Algorithms, Symbolic Algebra, Automatic Calculation}} (1999),
  \eprint{hep-ph/9905298}.

\bibitem{Ruijl:2017dtg}
B.~Ruijl, T.~Ueda and J.~Vermaseren,
\newblock \emph{{FORM version 4.2}}  (2017),
\newblock \eprint{1707.06453}.

\bibitem{Steinhauser:2000ry}
M.~Steinhauser,
\newblock \emph{{MATAD: A Program package for the computation of MAssive
  TADpoles}},
\newblock Comput. Phys. Commun. \textbf{134}, 335 (2001),
\newblock \doi{10.1016/S0010-4655(00)00204-6},
\newblock \eprint{hep-ph/0009029}.

\bibitem{Smirnov:2019qkx}
A.~V. Smirnov and F.~S. Chuharev,
\newblock \emph{{FIRE6: Feynman Integral REduction with Modular Arithmetic}},
\newblock Comput. Phys. Commun. \textbf{247}, 106877 (2020),
\newblock \doi{10.1016/j.cpc.2019.106877},
\newblock \eprint{1901.07808}.

\bibitem{Nogueira:1991ex}
P.~Nogueira,
\newblock \emph{{Automatic Feynman graph generation}},
\newblock J. Comput. Phys. \textbf{105}, 279 (1993),
\newblock \doi{10.1006/jcph.1993.1074}.

\bibitem{vanRitbergen:1998pn}
T.~van Ritbergen, A.~N. Schellekens and J.~A.~M. Vermaseren,
\newblock \emph{{Group theory factors for Feynman diagrams}},
\newblock Int. J. Mod. Phys. A \textbf{14}, 41 (1999),
\newblock \doi{10.1142/S0217751X99000038},
\newblock \eprint{hep-ph/9802376}.

\bibitem{Lee:2012cn}
R.~N. Lee,
\newblock \emph{{Presenting LiteRed: a tool for the Loop InTEgrals REDuction}}
  (2012),
\newblock \eprint{1212.2685}.

\bibitem{Lee:2013mka}
R.~N. Lee,
\newblock \emph{{LiteRed 1.4: a powerful tool for reduction of multiloop
  integrals}},
\newblock J. Phys. Conf. Ser. \textbf{523}, 012059 (2014),
\newblock \doi{10.1088/1742-6596/523/1/012059},
\newblock \eprint{1310.1145}.

\bibitem{Herren:2020ccq}
F.~Herren,
\newblock \emph{{Precision Calculations for Higgs Boson Physics at the LHC -
  Four-Loop Corrections to Gluon-Fusion Processes and Higgs Boson
  Pair-Production at NNLO}},
\newblock Ph.D. thesis, KIT, Karlsruhe,
\newblock \doi{10.5445/IR/1000125521} (2020).

\bibitem{Davies:2019esq}
J.~Davies, F.~Herren, G.~Mishima and M.~Steinhauser,
\newblock \emph{{NNLO real corrections to $gg\to HH$ in the large-$m_t$
  limit}},
\newblock PoS \textbf{RADCOR2019}, 022 (2019),
\newblock \doi{10.22323/1.375.0022},
\newblock \eprint{1912.01646}.

\bibitem{Davies:2019xzc}
J.~Davies, F.~Herren, G.~Mishima and M.~Steinhauser,
\newblock \emph{{Real-virtual corrections to Higgs boson pair production at
  NNLO: three closed top quark loops}},
\newblock JHEP \textbf{05}, 157 (2019),
\newblock \doi{10.1007/JHEP05(2019)157},
\newblock \eprint{1904.11998}.

\bibitem{Davies:2021kex}
J.~Davies, F.~Herren, G.~Mishima and M.~Steinhauser,
\newblock \emph{{Real corrections to Higgs boson pair production at NNLO in the
  large top quark mass limit}}  (2021),
\newblock \eprint{2110.03697}.

\bibitem{Davies:2019wmk}
J.~Davies, F.~Herren and M.~Steinhauser,
\newblock \emph{{Top Quark Mass Effects in Higgs Boson Production at Four-Loop
  Order: Virtual Corrections}},
\newblock Phys. Rev. Lett. \textbf{124}(11), 112002 (2020),
\newblock \doi{10.1103/PhysRevLett.124.112002},
\newblock \eprint{1911.10214}.

\bibitem{Davies:2021zbx}
J.~Davies and F.~Herren,
\newblock \emph{{Higgs boson decay into photons at four loops}},
\newblock Phys. Rev. D \textbf{104}(5), 053010 (2021),
\newblock \doi{10.1103/PhysRevD.104.053010},
\newblock \eprint{2104.12780}.

\bibitem{Grigo:2015dia}
J.~Grigo, J.~Hoff and M.~Steinhauser,
\newblock \emph{{Higgs boson pair production: top quark mass effects at NLO and
  NNLO}},
\newblock Nucl. Phys. B \textbf{900}, 412 (2015),
\newblock \doi{10.1016/j.nuclphysb.2015.09.012},
\newblock \eprint{1508.00909}.

\bibitem{Davies:2019djw}
J.~Davies and M.~Steinhauser,
\newblock \emph{{Three-loop form factors for Higgs boson pair production in the
  large top mass limit}},
\newblock JHEP \textbf{10}, 166 (2019),
\newblock \doi{10.1007/JHEP10(2019)166},
\newblock \eprint{1909.01361}.

\bibitem{deFlorian:2013jea}
D.~de~Florian and J.~Mazzitelli,
\newblock \emph{{Higgs Boson Pair Production at Next-to-Next-to-Leading Order
  in QCD}},
\newblock Phys. Rev. Lett. \textbf{111}, 201801 (2013),
\newblock \doi{10.1103/PhysRevLett.111.201801},
\newblock \eprint{1309.6594}.

\bibitem{Gehrmann:2010ue}
T.~Gehrmann, E.~W.~N. Glover, T.~Huber, N.~Ikizlerli and C.~Studerus,
\newblock \emph{{Calculation of the quark and gluon form factors to three loops
  in QCD}},
\newblock JHEP \textbf{06}, 094 (2010),
\newblock \doi{10.1007/JHEP06(2010)094},
\newblock \eprint{1004.3653}.

\bibitem{Actis:2008ts}
S.~Actis, G.~Passarino, C.~Sturm and S.~Uccirati,
\newblock \emph{{NNLO Computational Techniques: The Cases H ---\ensuremath{>}
  gamma gamma and H ---\ensuremath{>} g g}},
\newblock Nucl. Phys. B \textbf{811}, 182 (2009),
\newblock \doi{10.1016/j.nuclphysb.2008.11.024},
\newblock \eprint{0809.3667}.

\bibitem{Niggetiedt:2020sbf}
M.~Niggetiedt,
\newblock \emph{{Exact quark-mass dependence of the Higgs-photon form factor at
  three loops in QCD}},
\newblock JHEP \textbf{04}, 196 (2021),
\newblock \doi{10.1007/JHEP04(2021)196},
\newblock \eprint{2009.10556}.

\end{thebibliography}

\nolinenumbers

\end{document}